\documentclass{emulateapj}

\slugcomment{\scriptsize{ApJ accepted, Preprint typset using \LaTeX style}}
\shorttitle{DEAD ZONE FORMATION AND NONSTEADY HYPERACCRETION IN COLLAPSAR DISKS}
\shortauthors{MASADA ET AL.}
\begin{document}
\title{Dead Zone Formation and Nonsteady Hyperaccretion in Collapsar
  Disks: A Possible Origin of Short-Term Variability in The Prompt
  Emission of Gamma-Ray Bursts}

\author{Youhei Masada\altaffilmark{1,2}, Norita
  Kawanaka\altaffilmark{3},Takayoshi Sano\altaffilmark{4}, and
  Kazunari Shibata\altaffilmark{1}}

\altaffiltext{1}{Kwasan and Hida Observatories, Kyoto University, 
Kyoto 607-8471, Japan; masada@kusastro.kyoto-u.ac.jp}
\altaffiltext{2}{Department of Astronomy, Kyoto University, Kyoto 606-8502, Japan}
\altaffiltext{3}{Yukawa Institute for Theoretical Physics, Kyoto University, Kyoto 606-8502, Japan}
\altaffiltext{4}{Institute of Laser Engineering, Osaka University, Osaka 565-0871, Japan} 
\begin{abstract}
The central engine of gamma-ray bursts (GRBs) is believed to be a hot
and dense disk with hyperaccretion onto a few solar-mass black hole. 
We investigate where the magnetorotational instability (MRI) actively
operates in the hyperaccretion disk, which can cause angular
momentum transport in the disk. 
The inner region of hyperaccretion disks can be neutrino opaque,
and the energy- and momentum-transport by neutrinos could affect
the growth of the MRI significantly. 
Assuming reasonable disk models and a weak magnetic field $B \lesssim
10^{14}\ \rm{G}$, it is found that the MRI is strongly
suppressed by the neutrino viscosity in the inner region of
hyperaccretion disks.
On the other hand, the MRI can drive active MHD turbulence in the
outer neutrino-transparent region regardless of the field strength. 
This suggests that the baryonic matter is accumulated into the inner
dead zone where the MRI grows inactively and the angular momentum
transport is inefficient. 
When the dead zone gains a large amount of mass and becomes
gravitationally unstable, intense mass accretion onto the central 
black hole would occur episodically through the gravitational torque. 
This process can be a physical mechanism of the short-term variability 
in the prompt emission of GRBs. 
Finally, the origin of flaring activities in the X-ray afterglow is
predicted in the context of our episodic accretion scenario. 
\end{abstract}

\keywords{accretion, accretion disks --- black hole physics --- gamma
  rays : bursts --- instabilities : magnetic fields}
\section{Introduction}
Gamma-ray bursts (GRBs) are the most energetic event in the universe.
GRBs are generally considered to be powered by hyperaccretion of
the matter onto a stellar-mass black hole ($\sim2$--$ 3M_{\sun}$),
which is formed in the context of the ``collapsar'' scenario or merging
scenarios of compact objects (Narayan et al. 1992; Woosley 1993; Paczynski 1998; 
MacFadyen \& Woosley 1999; Ruffert \& Janka 1999, 2001; 
M\'esz\'aros 2002; Piran 2005). 
The hyperaccretion rate is of the order of $0.1$--$1\ M_\sun s^{-1}$
and the release of gravitational energy powers the burst.
The radiative energy ejected through relativistic jets is 
expected to account for the observed $\gamma$-ray emission. 

The key process for releasing the gravitational energy 
is angular momentum transport in the disk (Lee \& Ramirez-Ruiz 2006). 
As in the cases of the other astrophysical accretion disks, 
magnetic turbulence initiated and sustained by the
magnetorotational instability (MRI) is believed to play an essential
role for the angular momentum transport in hyperaccretion disks
(Balbus \& Hawley 1991; Lee et al. 2005; Fujimoto et al. 2006; Nagataki et al. 2006). 
Detailed linear and nonlinear analysis of the MRI
facilitate our understanding the role of the MRI in various accretion
systems (Hawley et al. 1995; Machida et al. 2000; Turner et al. 2002; 
De Villiers et al. 2003; Sano et al. 2004). 
However, the physical conditions of hyperaccretion disks are quite
different from the other accretion systems. 
Because the hyperaccretion disks are very dense and hot like supernova
cores, they can be cooled through the neutrino radiation 
(Popham et al. 1999; Narayan et al. 2001; Di Matteo et al. 2002). 
In addition, the energy and momentum are mainly transported by
neutrinos in the neutrino-opaque region. 
However, the detailed analysis of the MRI in hyperaccretion disks
including such microscopic effects has not been done yet except for
Araya-G\'ochez \& Vishniac (2004). 

Masada et al. (2007; hereafter MSS07) have investigated the effects of 
the neutrino transport on the MRI in proto-neutron stars (PNSs), 
and shown that the heat, chemical, and viscous diffusions caused by the 
neutrino transport have great impacts on the MRI. 
The heat and chemical diffusions can reduce the effect of
stratifications, and the neutrino viscosity suppresses the growth of
the MRI. Because the resistivity of hot nuclear matter is
relatively low (Yakovlev \& Shalybkov 1991; Miralles et al. 2002), 
the magnetic diffusion is much smaller than the other diffusion 
processes, and hardly affects the MRI inside PNSs. 

The hyperaccretion disk is considered to have similar physical
properties to the PNS. In particular, neutrinos can be trapped in the inner dense region 
of hyeraccretion disks, and the energy and momentum are mainly
transported by the neutrinos. Therefore, we can apply the results of MSS07 
to hyperaccretion disks with little changes. 
 
The growth time of the MRI in the absence of viscosity is given by
$\lambda / v_A$, where $\lambda$ is the wavelength of a perturbation
and $v_A = B/(4\pi\rho)^{1/2}$ is the Alfv\'en speed.
The growth of the MRI is suppressed dramatically
if the growth time is longer than the viscous damping time
$\sim \lambda^2/\nu$, where $\nu$ is the kinematic viscosity (MSS07). 
Then, a large enough viscosity can reduce the linear growth of the MRI. 
Clearly, as the wavelength becomes longer, larger viscosities are 
required to suppress the MRI. 
Because the typical wavelength of the MRI is $\lambda
\sim v_A/\Omega$,  the condition for the linear growth of the MRI can be
written as
\begin{equation}
Re \equiv \displaystyle{\frac{LU}{\nu}} =
\displaystyle{\frac{v_A^2}{\nu \Omega}} \gtrsim 1 \label{eq1} \;, 
\end{equation}
where $Re$ is the Reynolds number,  
and $\Omega$ is the angular velocity. Here we choose $v_A/\Omega$ 
as the typical length scale $L$ and $v_A$ as the typical velocity $U$. 
The Reynolds number is a good indicator for rapid growth of the
MRI in hyperaccretion disks. 
In the dense neutrino-opaque matter, the kinematic viscosity via the
neutrino transport would be quite large, so that the condition given
by equation~(\ref{eq1}) may not be satisfied.
Therefore, in this paper, we investigate where the MRI operates 
in hyperaccretion disks focusing on the neutrino viscous effect.

The plan of this paper is as follows: In \S~2, we construct 
reasonable disk models, and derive simple prescriptions for 
the neutrino opacity and viscosity. 
Linear growth of the MRI in hyperaccretion disks is studied in \S~3 
using the dispersion equation derived in MSS07. 
Based on the linear stability analysis of the MRI, 
we propose a new evolutionary scenario of collapsar disks that can
explain the short-term variability in long GRBs in \S~4. 
In \S~5, we discuss a possible origin of flaring activities 
in the X-ray afterglow, which have been discovered by the {\it{Swift}} satellite. 
Finally, we summarize our main findings in \S~6. 

\section{THE PHYSICAL MODELS}
\subsection{Neutrino Depth and Viscosity}
We need to estimate the amplitude of the kinematic viscosity in
hyperaccretion disks to investigate where the MRI is active.
The momentum transport by neutrinos plays an important role in the
neutrino-opaque regions. 
In contrast, its influences can be 
negligible when the accreting gas is transparent to the neutrino. 
We treat, for simplicity, the neutrino transport with the 
diffusion approximation. The kinematic viscosity is assumed to be
zero in the neutrino-transparent regions. 
Note that the diffusion approximation fails when 
the neutrino depth reaches to unity. 
However, we focus mainly on the qualitative features of the MRI 
in hyperaccretion disks, so that we adopt this simple treatment in
this paper. 

First we derive the neutrino depth and kinematic viscosity via the
neutrino transport. 
Both absorption and scattering processes of the neutrino 
contribute to the neutrino depth. 
Absorption onto the nucleons by the inverse of the URCA process
gives rise to a large portion of the absorptive neutrino depth. 
The other absorptive processes, such as the 
inverse of pair annihilation, bremsstrahlung, 
and plasmon decay would be negligible (Kohri \& Mineshige 2002). 
The absorptive neutrino depth is, thus, described by 
$\tau_{\rm abs}\simeq 4.5\times 10^{-39} T^2 X_{\rm nuc}\rho H$,
where $T$ is the temperature, $\rho$ is the density, 
$H$ is the scale height of the disk, and $X_{\rm nuc}$ is the mass
fraction of free-nucleons (Di Matteo et al. 2002). 
The scattering neutrino depth through elastic scattering of
background nucleons is given by 
$\tau_{\rm sc} = 3\rho\kappa_{\rm sc} H \simeq 8.1\times 10^{-39} T^2 \rho H$,
where the factor $3$ assumes equivalent contribution for all neutrino
species and $\kappa_{\rm sc}$ is the mass scattering coefficient
(Tubbs \& Schramm 1975; Shapiro \& Teukolsky 1983).  

Total neutrino depth $\tau_{\rm tot}$ is given by the sum of 
the absorptive and scattering neutrino depth, 
\begin{equation}
\tau_{\rm tot} = \tau_{\rm abs} + \tau_{\rm sc} 
= 1.3\times 10^{-38} T^2\rho H
\label{eq2} \;. 
\end{equation}
Here $X_{\rm nuc}$ can be assumed to be unity in the inner dense region 
of hyperaccretion disks. The diffusion approximation 
for the neutrino transport can be used in the region 
where the condition $\tau_{\rm tot} > 2/3$ is satisfied. In such
regions, 
the averaged mean-free-path of the neutrino is defined by 
$\langle\lambda\rangle = H/\tau_{\rm tot}$, 
and the neutrino viscosity $\nu$ is described by 
\begin{equation}
\nu = \displaystyle{\frac{4}{15}} 
\displaystyle{\frac{U_\nu}{\rho c}} 
\langle\lambda\rangle 
\approx 5.2 \times 10^{12} T^2 \rho^{-2} 
\ \ \rm{cm^2\ sec^{-1}} \label{eq3} \;,
\end{equation} 
where $U_\nu = 4\sigma_B T^4/c$ is the neutrino energy density 
(van den Horn \& van Weert 1984; Burrows \& Lattimer 1988). 
Here $\sigma_B$ and $c$ are the Stefan-Boltzmann constant 
and the speed of light, respectively. 
Spatial distributions of the neutrino depth and 
viscosity in hyperaccretion disks can be obtained from 
equations~(\ref{eq2}) and (\ref{eq3}) in so far as the disk structure
is determined. 

\subsection{Structure of Hyperaccretion Disks}
We adopt basic equations based on the Newtonian dynamics 
and assume a quasi-steady structure of hyperaccretion disks. 
Various steady disk models have been proposed as the central engine of
GRBs taking into account detailed microphysics and/or general
relativity (Popham et al. 1999; 
Narayan et al. 2001; Kohri \& Mineshige 2002; Di Matteo et al. 2002;
Kohri et al. 2005; Chen \& Beloborodov 2007; Kawanaka \& Mineshige 2007). 
These models are constructed in the framework of 
the $\alpha$-prescription of turbulent viscosity (Shakura \& Sunyaev 1973). 
However, we now have an interest in the features of the MRI 
that causes the turbulent viscosity. 
It is, thus, necessary to construct disk models
independent of the $\alpha$-parameter.
For simplicity, we adopt power-law models for the radial
distributions of all physical quantities.

The disk surface density $\Sigma(r)$ 
is one of the most important quantities in constructing the disk structure. 
We assume a power-law distribution with an index $q$, 
\begin{equation}
\Sigma (r) = \Sigma_0 \hat{r}^{-q}  \;, \label{eq4} 
\end{equation}
where $\hat{r} = r/r_s$ is the distance from the central black hole
normalized by the Schwarzschild radius.
The Schwarzschild radius is given by $r_s = 2GM_{\rm BH}/c^2 =
8.9\times10^5 M_3$ cm, where $M_3 = M_{\rm BH} / (3 M_\sun)$ is a
mass of black hole $M_{\rm BH}$ normalized by 3 $M_{\sun}$. 
Here $\Sigma_0$ is a reference value of the surface density at $r = r_s$. 
When the baryonic matter of its mass $\sim 0.1$--$1\ M_\sun$ is
distributed within $\sim 10^{7-8}\ \rm{cm}$ from the central black
hole, the averaged surface density is in a range of $\sim
10^{17-18}\ \rm{g\ cm^{-2}}$ (Woosley 1993). 
Thus, we choose the reference value as 
$\Sigma_0 = 1.0\times 10^{18} f_\Sigma \ \rm{g\ cm^{-2}}$, 
where $f_\Sigma $ is an arbitrary parameter. 

Another key quantity is the disk temperature, which is assumed to
retain a power-law distribution with an index $p$, 
\begin{equation}
T(r) = T_0 \hat{r}^{-p}  \;, \label{eq5} 
\end{equation}
where $T_0$ is a reference value at $r = r_s$. 
A much less extreme assumption would be that the thermal energy is
$\sim 1$ -- $10$ \% of the gravitational energy of the disk
($\sim GM_{\rm BH}/r$). 
Then the typical disk temperature 
is estimated as $\sim 10^{11-12}\ \rm{K} $ at $r = r_s$. 
The reference value is chosen as $T_0 = 4.3\times 10^{11} f_T \ \rm{K}$, 
where $f_T$ is another arbitrary parameter. 
This temperature is equivalent to the Fermi energy of the partially 
degenerate electrons (Lee et al. 2005). 

Assuming the disk in which the gas pressure dominates the 
degenerate and radiation pressure, the sound speed is given by 
\begin{equation}
c_s = \left(\displaystyle{\frac{k_B T}{m_p}}\right)^{1/2} = 6.1
\times 10^9 f_T^{1/2} \hat{r}^{-p/2}\ \ \rm{cm\ sec^{-1}} \;,
\label{eq6}
\end{equation}
where $k_B$ is the Boltzmann constant and $m_p$ is the proton mass. 
In a gravitationally stable disk, the vertical component of the
gravity in the disk is contributed by the central black hole. 
The hydrostatic equilibrium in the vertical direction
determines the scale height of the disk, 
\begin{equation}
H = \displaystyle{\frac{c_s}{\Omega}} = 2.5\times 10^5 f_T^{1/2}M_3
\hat{r}^{-(p-3)/2}\ \ \rm{cm} \;, \label{eq7}
\end{equation}
where the disk is assumed to rotate with the Keplerian angular
velocity, $\Omega = \Omega_K = 2.4\times 10^4 M_3^{-1}\hat{r}^{-3/2}\
\rm{sec^{-1}}$. 
Then the density structure can be evaluated as
\begin{equation}
\rho = \displaystyle{\frac{\Sigma}{2 H}} = 2.0 \times 10^{12}
f_\Sigma f_T^{-1/2} M_3^{-1} 
\hat{r}^{(p-2q-3)/2} \ \ \rm{g\ cm^{-3}} \;. \label{eq8}
\end{equation} 
Using equations (\ref{eq5}), (\ref{eq7}), and (\ref{eq8}),
the radial profiles of the neutrino depth and 
viscosity are given by
\begin{equation}
\tau_{\rm tot}  =  1.3\times 10^3 f_\Sigma f_T^2 \hat{r}^{-2p-q} \;,
\label{eq9} 
\end{equation}
and 
\begin{equation}
\nu  = 8.1 \times 10^{10} f_\Sigma^{-2} f_T^3 M_3^{-1}
\hat{r}^{3(1-p)+2q} \ \ \rm{cm^2\ sec^{-1}} \;. \label{eq10}
\end{equation}

The strength of the magnetic field is also an important quantity to
determine the growth of the MRI in hyperaccretion disks, although it is
highly uncertain in the context of GRBs. 
We assume that the pre-collapse core of a massive star has 
a uniform magnetic field with the same amplitude as strongly magnetized
white dwarfs, that is $B_p \approx 10^{9}$ G and has an average density of 
$\rho_{\rm core}  \approx 2 \times 10^{9}\ {\rm g\ cm}^{-3}$. 
Considering that the magnetic flux is conserved in the core-collapse
phase ($B \propto \rho^{2/3}$),
the radial structure of the poloidal magnetic field in hyperaccretion
disks is obtained as
\begin{equation}
B_{p,{\rm disk}} = 1.0 \times 10^{11} 
f_\Sigma^{2/3} f_T^{-1/3} f_B M_3^{-2/3}
\hat{r}^{(p-2q-3)/3} \ \ \rm{G} \;, \label{eq11}
\end{equation}
where $f_B$ is the arbitrary magnetic parameter. 
The Alfv\'en speed in the disk is then given by 
\begin{equation}
v_A = 2.0 \times 10^4 f_\Sigma^{1/6} f_T^{-1/12} f_B M_3^{-1/6}
\hat{r}^{(p-2q-3)/12}\ \rm{cm\ sec^{-1}} \;. \label{eq12}
\end{equation}

The power-law indexes of the surface density and temperature are
determined by the thermal equilibrium in the disk.
There are two main cooling processes in hyperaccretion disks, which
are the advection cooling and neutrino one. 
Assuming that the advection cooling 
dominates the neutrino one, the power-law indexes of the surface density and 
temperature are given by $p=1.0$ and $q=0.5$ 
(Di Matteo et al. 2002; Chen \& Beloborodov 2007).
This corresponds to the ADAF (advection dominated accretion flow) type disk.
We can also choose the power-law indexes as $p=1.5$ and $q=0.0$,
if the neutrino cooling is dominant.
This is so called the NDAF (neutrino-cooling dominated accretion flow)
type disk, but the neutrino is assumed to be thermal here (Woosley 1993).
When the inner dense region is opaque to the neutrino, 
the disk structure would approach either of these two models. 
On the other hand, the disk structure may be quite different from
them in the neutrino-transparent region (Popham et al. 1999; Narayan
et al. 2001). However, since our main interest is in the
neutrino-opaque region, the disk models with single power-law
distributions are adopted in this paper.

The linear stability analysis shows that a Keplerian disk 
is gravitationally stable only when Toomre's $Q \equiv \Omega c_{\rm s}/(\pi
G \Sigma)$ is larger than about unity
(Toomre 1964; Goldreich \& Lynden-Bell 1965). The disk model given by
equations~(\ref{eq4}) -- (\ref{eq8}) 
is gravitationally stable when the inequality
\begin{equation}
Q = 7.0 \times 10^2 f_\Sigma^{-1}f_{\rm T}^{1/2}M_3^{-1}\hat{r}^{-(p-2q+3)/2} \gtrsim 1 \;, \label{eq13}
\end{equation}
is fulfilled. Since we are concerned with gravitationally stable disks,
we need to choose appropriate values for $f_\Sigma$ and $f_{\rm T}$
satisfying the condition given by equation (\ref{eq13}). 
Assuming the baryonic mass of hyperaccretion disks 
is $M_{\rm disk} \simeq 1 M_{\sun}$, the outermost radius of the
disk is about $70 r_{\rm s}$ for the ADAF-type disk and $30 r_s$ for
the NDAF-type disk. 
Then, in order that the entire disk is gravitationally stable, 
these parameters should be 
$f_{\Sigma}^{-1} f_{\rm T}^{1/2} \gtrsim 1$ 
and 
$f_{\Sigma}^{-1} f_{\rm T}^{1/2} \gtrsim 3$ 
for the ADAF-type and NDAF-type disks, respectively.
We consider the evolution of the disks with gravitationally unstable
outer regions in \S~6.

\section{Linear Growth of the MRI in Hyperaccretion Disks} 
We evaluate the Reynolds number defined by equation~(\ref{eq1}) for
hyperaccretion disks using equations~(\ref{eq10}) and (\ref{eq12}).
The radial profile of the Reynolds number is given by  
\begin{equation}
Re = 2.1\times 10^{-7} f_\Sigma^{7/3} f_T^{-19/6} f_B^2 M_3^{5/3} 
\hat{r}^{(19p -14q-12)/6} \;. \label{eq14}
\end{equation}
The Reynolds number is independent of the radius
$\hat{r}$ for the ADAF models.
For the NDAF models, on the other hand, the Reynolds number increases
as $\hat{r}$ increases.
Then the neutrino viscosity is more effective in the inner part of the
disks. 

The condition that the growth rate of the MRI takes of the order of
the angular velocity is $Re \gtrsim 1$.
For the case of $f_\Sigma = f_T = 1$ (our fiducial model), it requires
that the magnetic field parameter should be $f_B \gtrsim 10^3$. 
This indicates that the strong magnetic field more than 
$10^{14} \rm{G}$ is necessary for rapid growth of the MRI in
hyperaccretion disks. 
When the magnetic field is weaker than this critical value, 
the inner region of the disk could be a ``dead zone'' where the
neutrino viscosity suppresses the MRI. 
To show this picture more concretely, in the following sections,
we solve the dispersion equation including the
neutrino transport,
and derive the maximum growth rate of the MRI in hyperaccretion disks.

\subsection{Dispersion Equation with the Neutrino Viscosity}
MSS07 derive the dispersion equation for the MRI including the effects
of the neutrino transport.
We apply it to find out the most unstable modes of the MRI in
hyperaccretion disks. 

The effects of the stratification due to the thermal and leptonic
gradients can be ignored in the disk.
We focus on the linear growth rate of the axisymmetric MRI.
This is because
the axisymmetric mode is the fastest growing mode of the MRI, and
the stability conditions for the axisymmetric and nonaxisymmetric MRI
are identical (MSS07).
The toroidal component of the field does not affect the linear and
nonlinear growth of the MRI if the field strength is subthermal (Blaes
\& Balbus 1994; Sano \& Stone 2002), so that only the poloidal field
is considered in our dispersion equation.
Ohmic dissipation would be negligible in hyperaccretion disks.

Assuming a uniform vertical field and considering only the
damping effect by the neutrino viscosity, the dispersion equation in
MSS07 can be written as 
\begin{equation}
\gamma^4 + a_3 \gamma^3 + a_2 \gamma^2 + a_1 \gamma + a_0 = 0 \;,
\label{eq15}
\end{equation}
where
$$
a_3 = 2\nu k^2 \;,\ \ \ a_2 = \nu^2 k^4 +2(k_z v_{A})^2 + (k_z/k)^2
\kappa^2 \;, 
$$
$$
a_1 = 2\nu k^2 (k_z v_A)^2 \;,\ \ \ a_0 = (k_z v_A)^4 -4(k_z/k)^2(k_z
v_A)^2\Omega^2 \;.
$$
Here $\gamma$ is the growth rate, $\kappa$ is the epicyclic frequency,
$\nu$ is the neutrino viscosity, and $k = (k_r^2 + k_z^2)^{1/2}$ is
the wavenumber. Radial and vertical wavenumbers are described by $k_r$
and $k_z$. 
For the Keplerian disks, the epicyclic frequency is equal to the
angular velocity.
Note that the dispersion equation (\ref{eq14}) is characterized by a
single parameter, which is the Reynolds number $Re = v_A^2/(\nu
\Omega)$.
We solve this dispersion equation numerically and show 
the maximum growth rate of the MRI in hyperaccretion disks. 

\subsection{The Maximum Growth Rates of MRI}
\begin{figure}
\begin{center}
\begin{tabular}{c}
\scalebox{0.6}{\rotatebox{0}{\includegraphics{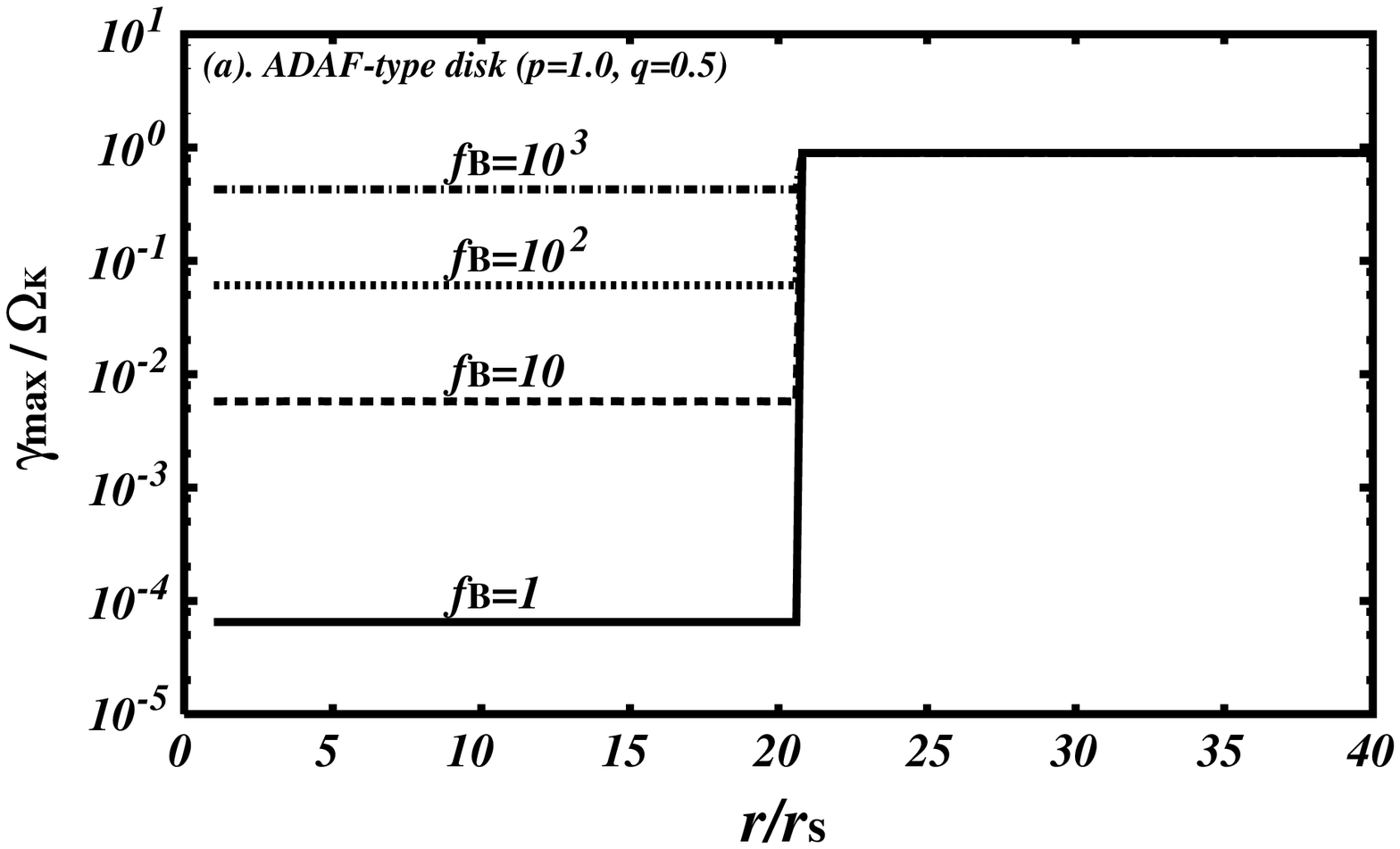}}} \\
\end{tabular}
\begin{tabular}{c}
\scalebox{0.6}{\rotatebox{0}{\includegraphics{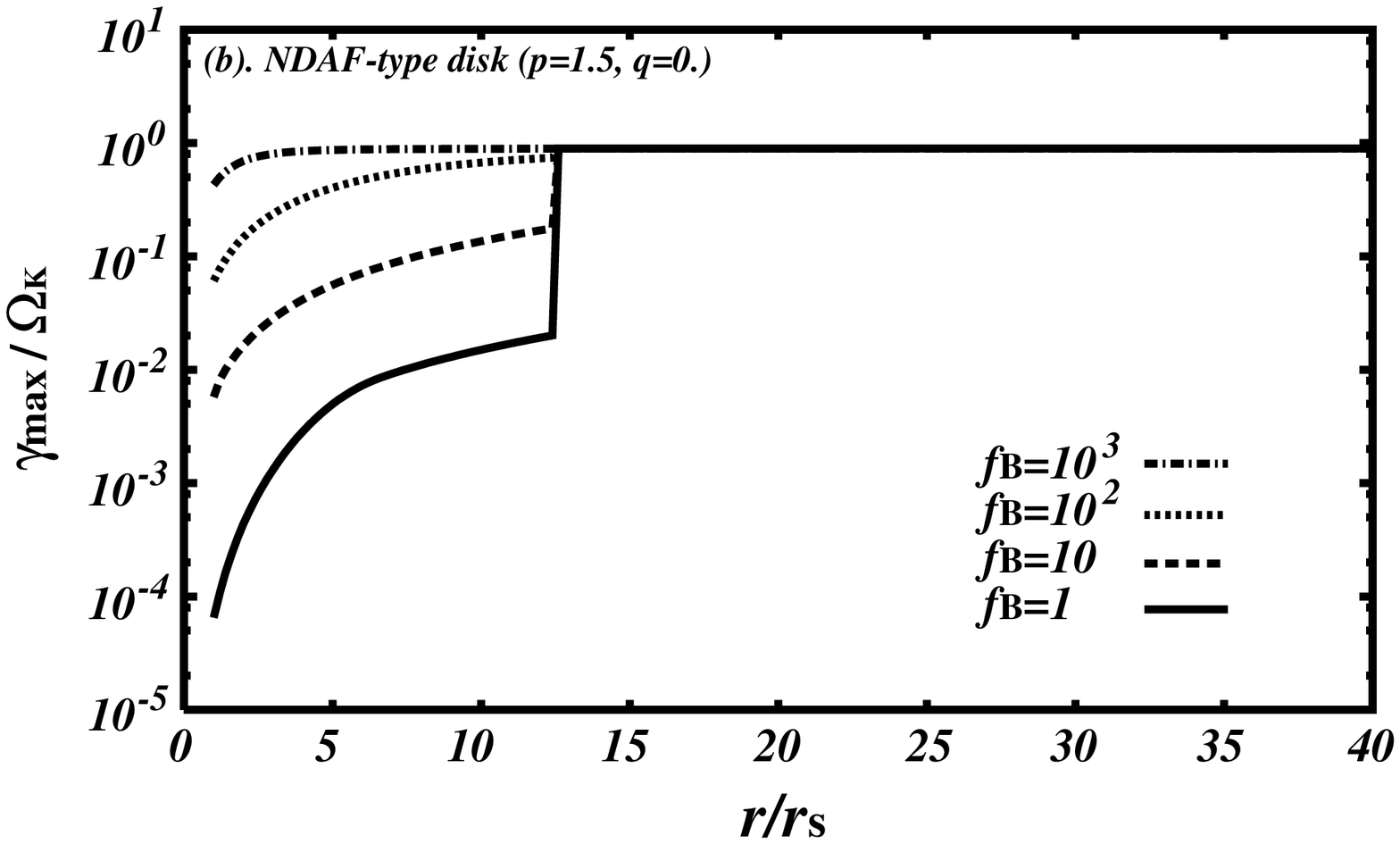}}} \\
\end{tabular}
\caption{The maximum growth rate of the MRI as a function of the disk
radius for the cases with different magnetic parameters $f_B = 1$, 10,
10$^2$, and $10^3$. Other arbitrary parameters 
are fixed to be unity in this figure, that is, $f_\Sigma = f_T = 1$. 
The vertical and horizontal axes are normalized by the 
Keplerian angular velocity $\Omega_K = 2.4\times 10^4 M_3^{-1}
\hat{r}^{-3/2}\ \rm{sec^{-1}}$ 
and the Schwarzschild radius $r_s=8.9\times 10^5 M_3 \ \rm{cm}$,
respectively. Here the mass of the central black hole is assumed to be
$M_{\rm BH} = 3M_3 M_\sun $.
Upper panel represents the case of the ADAF-type disk ($p=1.0,q=0.5$), 
and lower panel is the case of the NDAF-type disk ($p=1.5,q=0.0$). }
\label{fig1}
\end{center}
\end{figure}

Figure~\ref{fig1} shows the maximum growth rate of the MRI as a
function of the disk radius for the cases with different magnetic
parameters $f_B =$ 1, 10, 10$^2$ and $10^3$. Upper and lower panels
indicate the results for the ADAF-type disk ($p=1.0$, $q=0.5$) and the
NDAF-type disk ($p=1.5$, $q=0.0$), respectively. 
The vertical and horizontal axes are normalized by 
the Keplerian angular velocity $\Omega_K$ and the Schwarzschild radius
$r_s$. 
The parameters $f_\Sigma$ and $f_T$ are assumed to be 
unity in this figure (the fiducial model). The critical radius
dividing the neutrino-opaque and neutrino-transparent regions locates
at $r_{\rm crit} \approx 20r_s$ for the ADAF-type disk and 
$r_{\rm crit} \approx 13r_s$ for the NDAF-type disk. 

The maximum growth rate of the MRI in the neutrino-opaque region
is much smaller than that in the neutrino-transparent region
if the magnetic parameter $f_B$ is smaller than $10^3$.
That is, the MRI is strongly suppressed by the neutrino viscosity 
when the magnetic field is weaker than the critical value, 
$B_{\rm crit} \approx 10^{14}$ G. 
Turbulent motions at the nonlinear stage cannot be sustained when the
growth rate of the MRI is much less than the angular velocity $\Omega$
(Sano \& Stone 2002).
In contrast, MHD turbulence driven by the MRI can grow actively in the
neutrino-transparent region regardless 
of the field strength, because the viscosity effect can be
neglected there.

\begin{figure}
\begin{center}
\begin{tabular}{c}
\scalebox{0.55}{\rotatebox{0}{\includegraphics{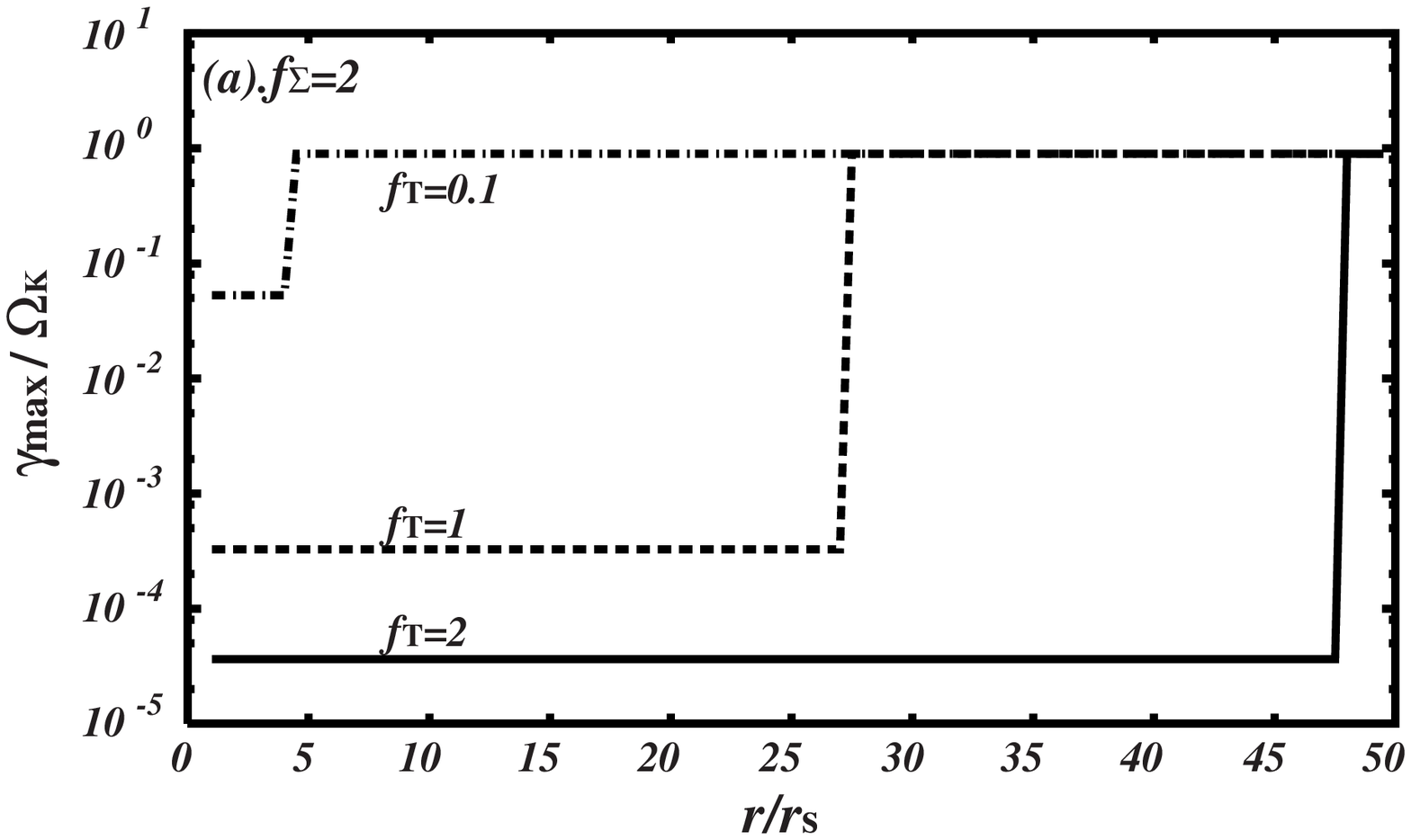}}} \\
\end{tabular}
\begin{tabular}{c}
\scalebox{0.55}{\rotatebox{0}{\includegraphics{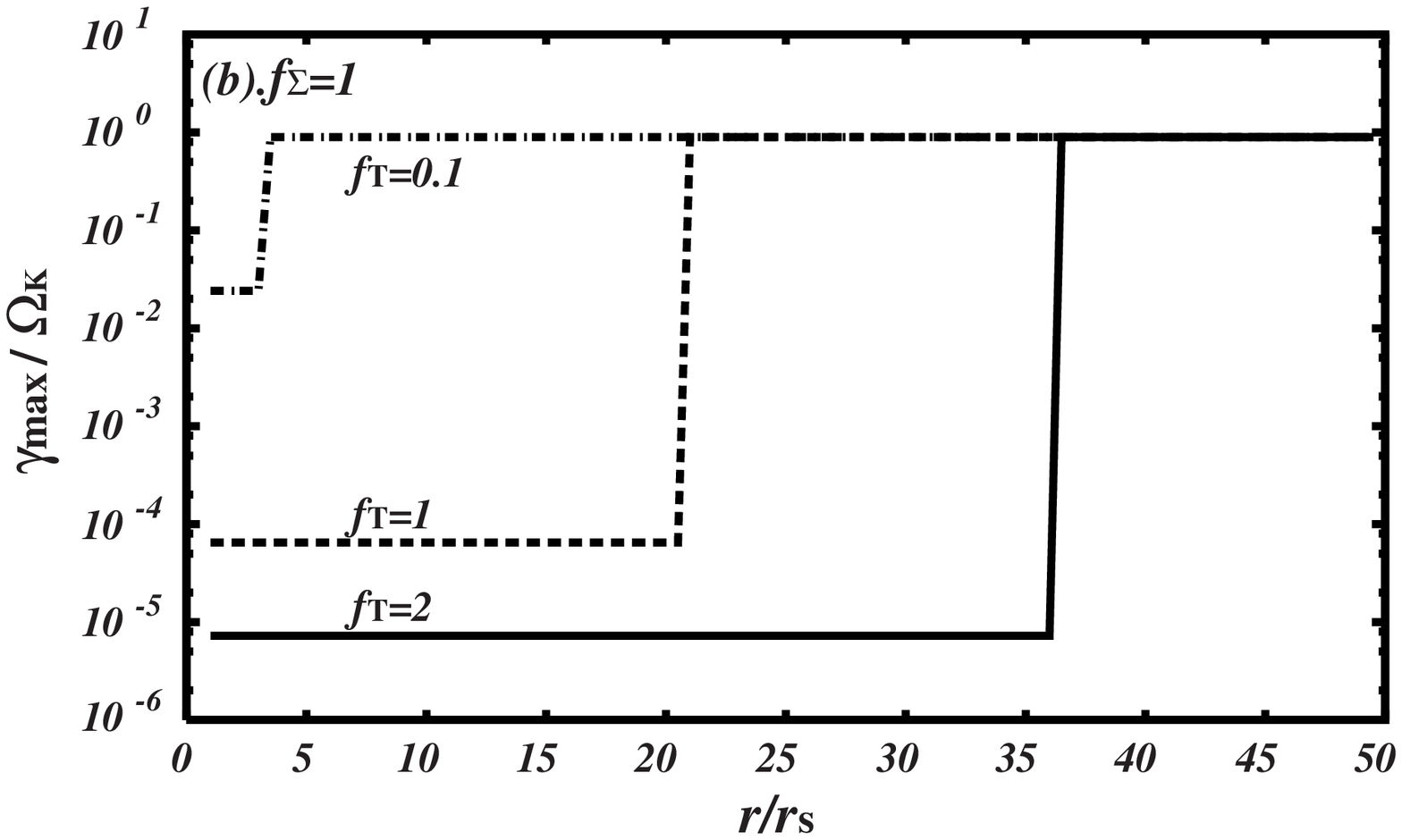}}} \\
\end{tabular}
\begin{tabular}{c}
\scalebox{0.55}{\rotatebox{0}{\includegraphics{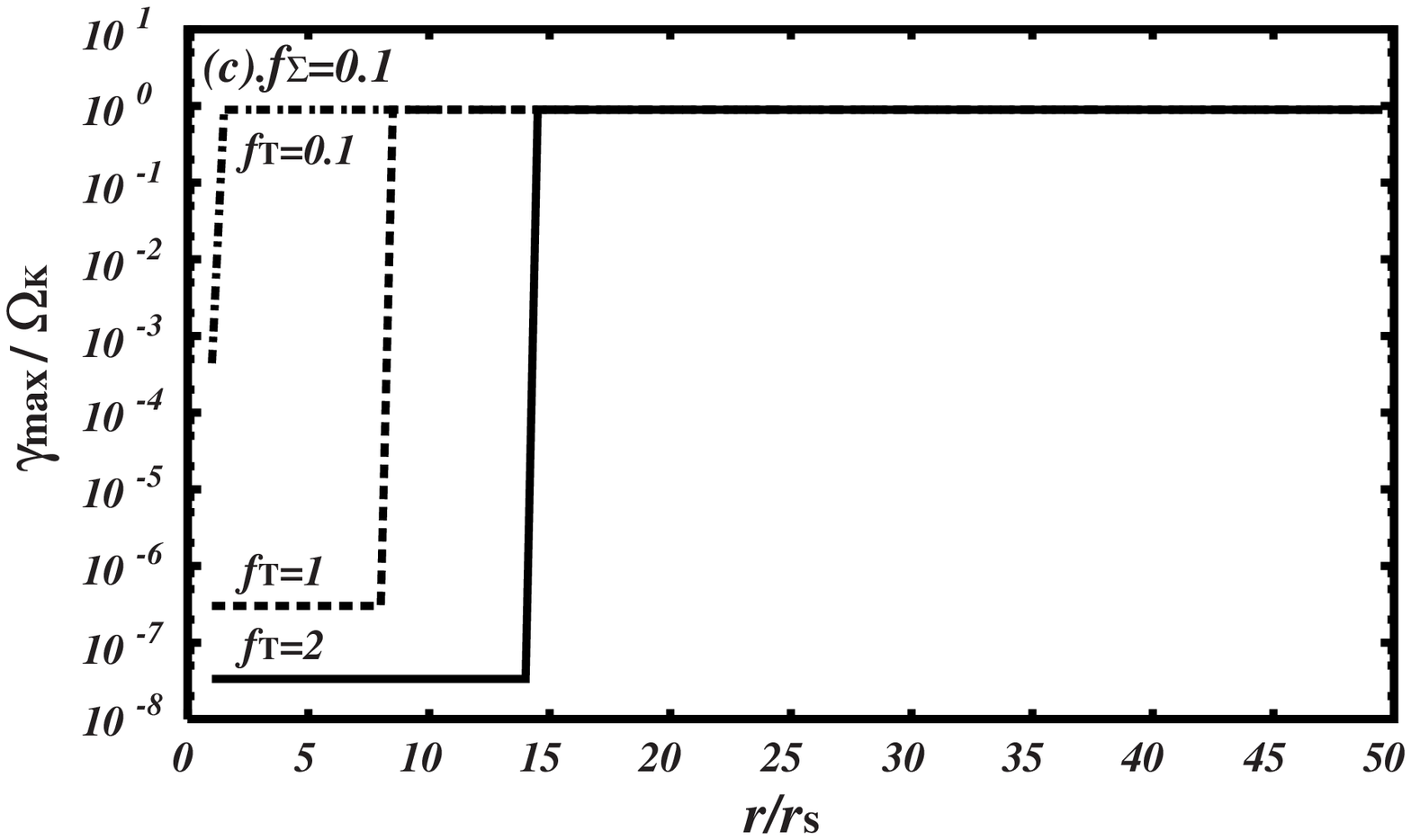}}} \\
\end{tabular}
\caption{The maximum growth rate of the MRI as a function of the disk
radius for the ADAF-type disk with different model parameters
$f_\Sigma$ and $f_T$.
Normalizations of the vertical and horizontal axes are the same as in
Figure~\ref{fig1}. 
The magnetic parameter $f_B$ is fixed to be unity. Upper panel
shows the $f_T$-dependence of the 
maximum growth rate for the case of $f_\Sigma = 2$. Middle and lower
panels are those for the cases of 
$f_\Sigma = 1$ and $f_\Sigma =0.1$, respectively. }
\label{fig2}
\end{center}
\end{figure}

\begin{figure}
\begin{center}
\begin{tabular}{c}
\scalebox{0.55}{\rotatebox{0}{\includegraphics{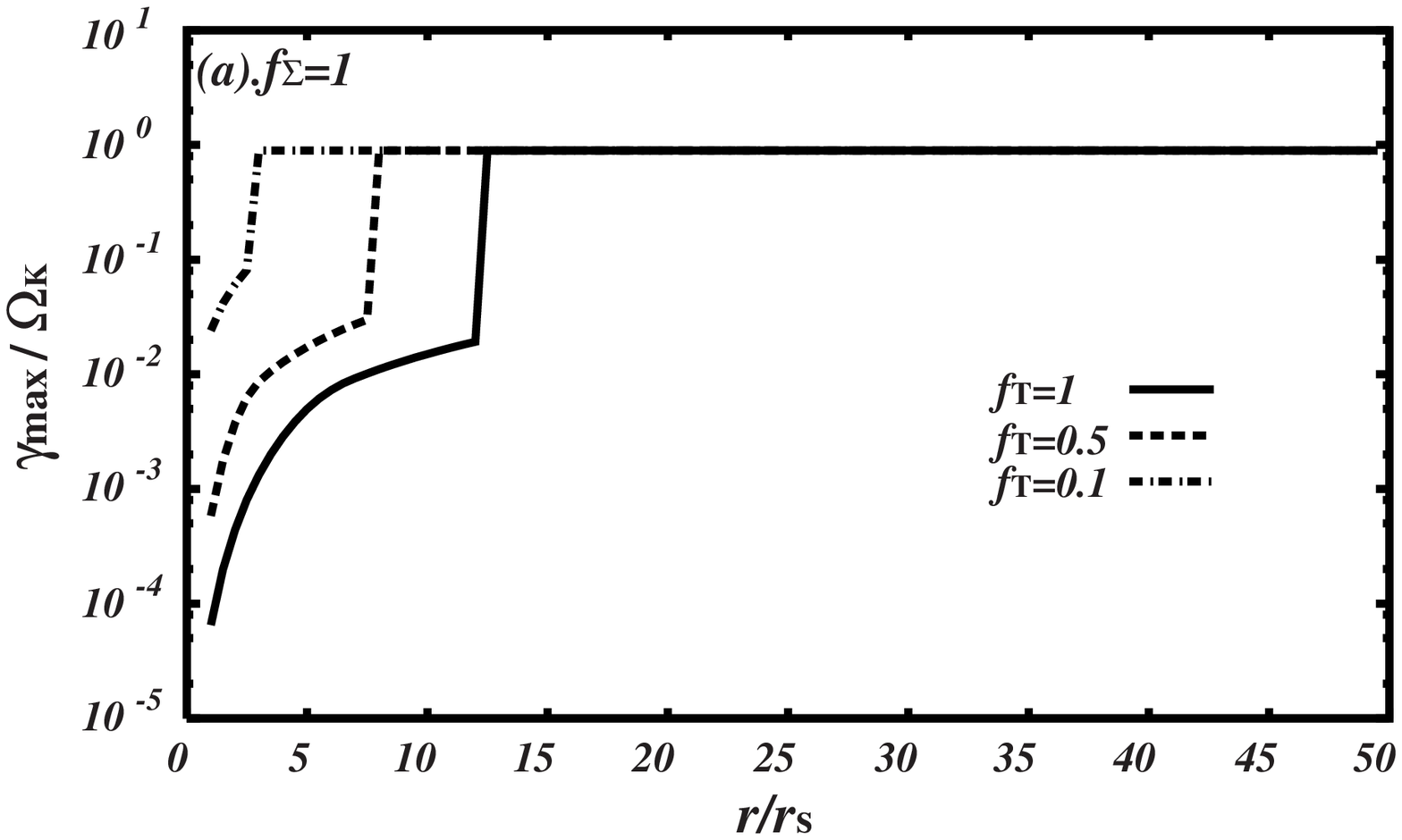}}} \\
\end{tabular}
\begin{tabular}{c}
\scalebox{0.55}{\rotatebox{0}{\includegraphics{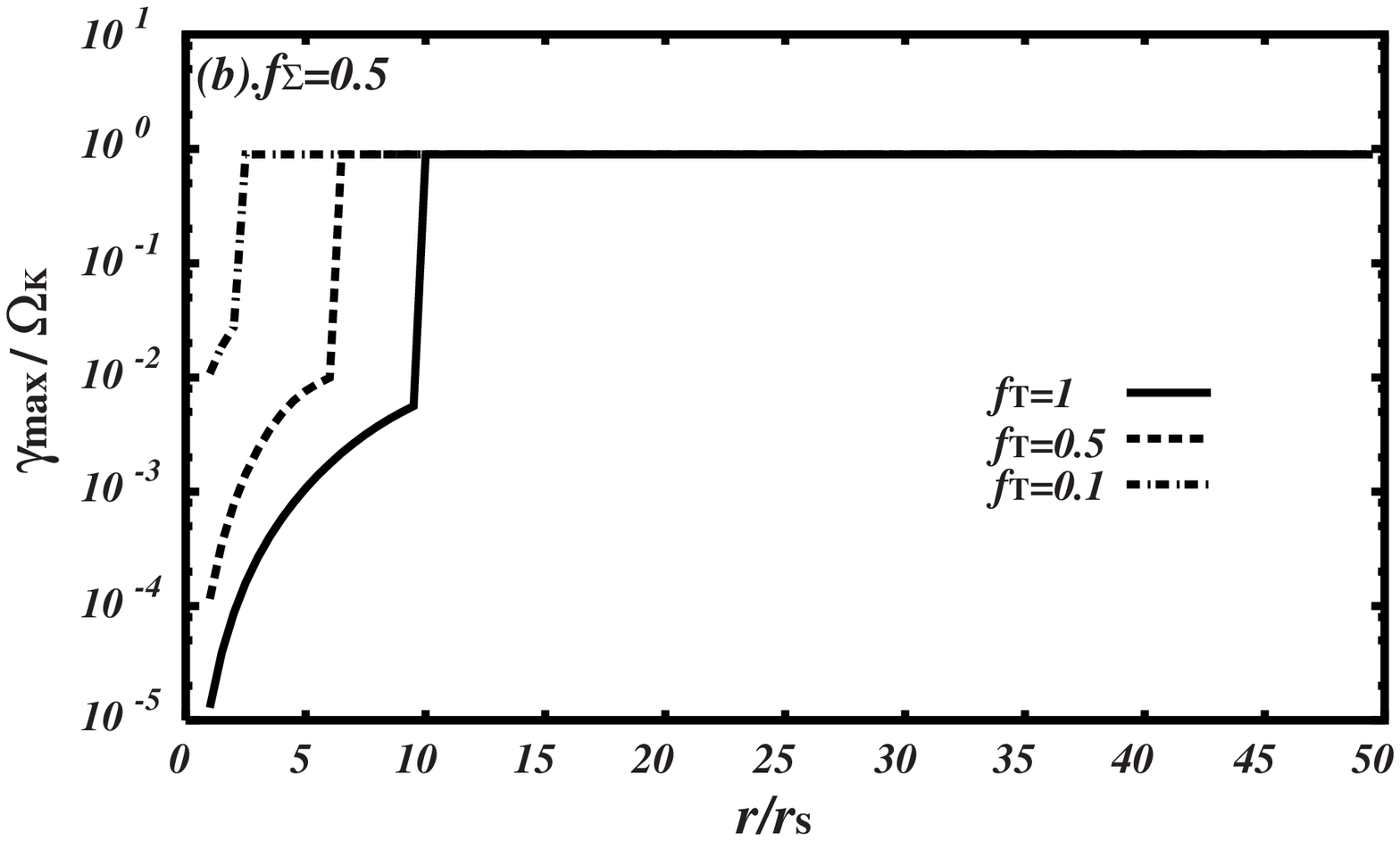}}} \\
\end{tabular}
\begin{tabular}{c}
\scalebox{0.55}{\rotatebox{0}{\includegraphics{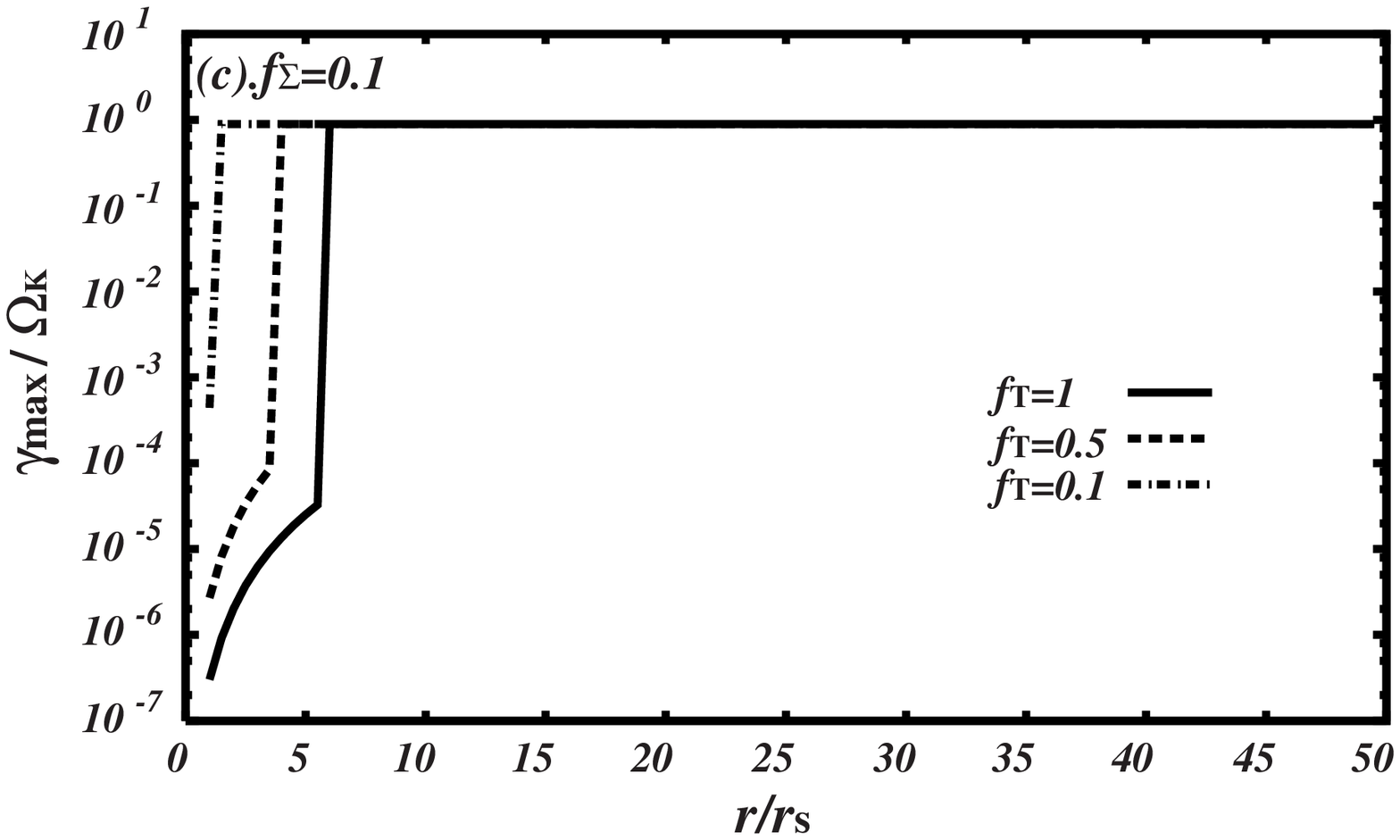}}} \\
\end{tabular}
\caption{The maximum growth rate of the MRI as a function of the disk
radius for the NDAF-type disk with different model parameters
$f_\Sigma$ and $f_T$.
Normalizations of the vertical and horizontal axes are the same as in
Figure~\ref{fig1}. 
The magnetic parameter $f_B$ is fixed to be unity. Upper panel
shows the $f_T$-dependence of the maximum growth rate for the case of 
$f_\Sigma = 2$. Middle and lower panels are those for the cases of
$f_\Sigma = 1$ and $f_\Sigma =0.1$, respectively.}
\label{fig3}
\end{center}
\end{figure}

The characteristics of the MRI in hyperaccretion disks depend on the
parameters $f_\Sigma$ and $f_T$. 
The maximum growth rate of the MRI for the cases with the different
disk temperature and surface density are investigated for the two
types of disk models in Figures~\ref{fig2} and \ref{fig3}. 
Normalizations of the vertical and horizontal axes are the same as in
Figure~\ref{fig1}. The magnetic parameters $f_B$ 
is assumed to be unity in these figures. 

For the both disk models, 
it is found that the maximum growth rate of the MRI decreases and the
size of the dead zone expands with increase of the disk temperature.
This is because the neutrino depth and viscosity increase 
with the disk temperature [see eqs.~(\ref{eq2}) and (\ref{eq3})]. 
On the other hand, the decrease of the surface density yields the
shrinking of the dead zone and decrease of the maximum growth rate. 
The contraction of the dead zone arises from the decrease of the
neutrino depth. 
The reduction of the maximum growth rate is caused by the rise of 
the neutrino viscosity with decrease of the surface density. 

These results indicate that the dead zones with the size of $\sim 10$
-- $20 r_s$ are accompanied with hot and dense hyperaccretion disks
($f_\Sigma \sim f_T \sim 1$). 
In what follows, we develop an evolutionary scenario of
hyperaccretion disks with the relatively large size of dead zone. 

\section{Episodic Accretion Model for GRBs}
As is described in the previous section, the MRI is 
suppressed by the neutrino viscosity in the neutrino-opaque region 
when the magnetic field is weaker than the critical value, $B_{\rm
  crit} \approx 10^{14} \rm{G}$. 
In contrast, the neutrino-transparent region (hereafter active region) 
is unstable for the MRI and its growth rate is of the order of the 
angular velocity. These features of the MRI in hyperaccretion disks 
have not been considered in previous work at all. 
However, this can be an important physical clue to reveal 
the central engine of GRBs. 

The existence of the dead zone could cause episodic 
hyperaccretion, which can be the source of the short-term variability
in the prompt emission of GRBs. We focus on the hyperaccretion disk 
formed in the context of ``collapsar'' scenario (Woosley 1993). 
In this scenario, the central $\sim 3 M_\sun$ of a massive star
collapses directly to a black hole, 
while infalling material with large specific angular momentum 
forms an accretion disk of its mass $\sim 1$ -- $2 M_\sun$. 

\subsection{Evolutionary Scenario of the Disks with Dead Zones}
We consider a relatively large dead zone ($r_{\rm crit} \sim 20 r_s$)
formed at the inner part of a collapsar disk. 
In such the case, angular momentum transport in the dead zone would be 
taken by the neutrino viscosity itself, not by the turbulent viscosity
sustained by the MRI. 
The size of the neutrino viscosity can be described by using the
$\alpha$-parameter, 
\begin{equation}
\alpha_\nu = \frac{\nu\Omega_K}{c_s^2} 
= 5.2\times 10^{-5} f_\Sigma^2 f_T^2 M_3^{-2} 
\hat{r}^{-2(p-q)+3/2}  \label{eq16} \;.
\end{equation}
Thus, the $\alpha$-value of the dead zone is typically $\sim 10^{-4}$
for the fiducial model ($f_\Sigma = \ f_T = 1$). 
The mass accretion rate, $\dot{M}_{\rm out}$,
onto the central black hole is given by 
\begin{equation}
\dot{M}_{\rm out} = 4 \pi r \rho H v_r \simeq 7.2 \times 10^{-4} \left(
\frac{\alpha_{\nu}}{10^{-4}} \right) f_\Sigma f_T M_3 \dot{M}_\sun 
\;, \label{eq17} 
\end{equation}
where the drift velocity is assumed to be $v_r = 3 \nu / (2 r)$ and
$\dot{M}_\sun$ is the mass accretion rate of $1 M_\sun \sec^{-1}$. 
This is the mass of the baryonic matter extracted from the dead zone
per unit time. 

On the other hand, in the active region, the turbulent viscosity
driven by the MRI would play an essential role in the angular
momentum transport. 
It is expected from various nonlinear simulations of the MRI that the
$\alpha$-parameter due to the turbulent viscosity could be
$\alpha_t \sim 10^{-2}$ or less
(Hawley et al. 1995; Stone et al. 1996; Sano et al. 2004). 
Therefore, the mass inflow rate into the dead zone $\dot{M}_{\rm in}$
is about
\begin{equation}
\dot{M}_{\rm in} \simeq 7.2 \times 10^{-2} \left(
\frac{\alpha_t}{10^{-2}} \right) f_\Sigma f_T M_3 \dot{M}_\sun 
\; \label{eq18} 
\end{equation}
This is the mass of the baryonic matter which flows into the dead zone
per unit time. 

From the equations~(\ref{eq17}) and (\ref{eq18}), it is expected that 
the baryonic matter is accumulated into the dead zone. 
The mass accumulation rate into the dead zone $\dot{M}_{\rm accu}$
is equivalent to the mass inflow rate, because the extracted mass from
the dead zone is negligible (i.e., $\dot{M}_{\rm in} \gg \dot{M}_{\rm
  out}$). 
If the mass accumulation with a rate $\dot{M}_{\rm accu} \approx
\dot{M}_{\rm in}$ continues, the inner dead zone becomes
gravitationally unstable 
at some stage. Then the gravitational torque may cause intermittent mass
accretion and drive discrete jets, which is thought to be the origin of 
short-term variability in the prompt emission of GRBs 
(Rees \& M\'esz\'aros 1994; Kobayashi et al. 1997). 

\begin{figure}
\begin{center}
\scalebox{0.6}{\rotatebox{0}{\includegraphics{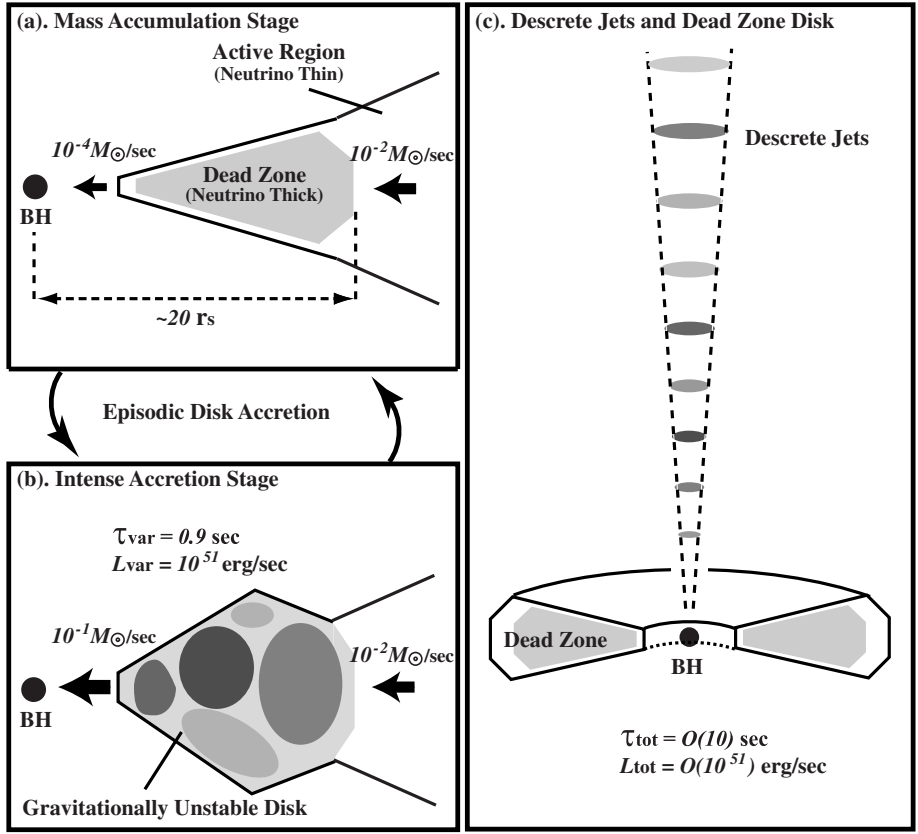}}}
\caption{
Schematic pictures of typical three evolutionary stages of a collapsar
disk with a dead zone. Panel~(a) shows the mass accumulation stage. In
this stage, the MRI is suppressed by the neutrino viscosity and angular
momentum transport is inefficient in the inner neutrino-opaque region.
On the other hand, in the outer neutrino-transparent region, the MRI can
drive MHD turbulence actively and the angular momentum is transported
efficiently. Then the baryonic matter is accumulated into the dead zone
during this stage. Panel~(b) represents the intense accretion stage. If the
baryonic matter is continually accumulated into the dead zone, it should
become gravitationally unstable at some evolutionary stage and the
intense mass accretion is triggered by the gravitational torque. 
The intense accretion would be terminated when the mass of the dead
zone becomes small enough to make the region stable for the
gravitational instability.
However, the mass accretion from the outer region continues, and thus
the quiescent dead zone approaches guradually to the intense accretion
stage again. Therefore, the dead zone region alternates between these
two stages and the explosive energy release occurs intermittently until the baryonic
matter of the outer active region is exhausted. Panel~(c) depicts the final stage 
after the episodic mass accretion is terminated. 
The dead zone disk with a low mass accretion rate will be left around the central black hole. 
Multiple discrete jets originated from the episodic mass accretion 
propagate with a relativistic speed. These multiple relativistic shells might be an origin of
the short-term variability in the prompt emission of GRBs.
}
\label{fig4}
\end{center}
\end{figure}

The prospective evolutionary scenario of collapsar disks with dead
zones is depicted schematically in Figure~\ref{fig4}. 
The dead zone region undergoes the quiescent phase and burst phase
repeatedly. 
At the quiescent phase (Fig. \ref{fig4}$a$), the neutrino viscosity is
the only mechanism of angular momentum transport.
The mass accretion rate onto the black hole is relatively small and the
mass of the dead zone increases during this phase.
When the gravitational instability set in, the gravitational torque
would enhance the accretion rate more than two orders of magnitude
(Fig. \ref{fig4}$b$).
The gravitational torque works until the dead zone region becomes
gravitationally stable as a result of the intense mass accretion onto
the central black hole.

The evolution of accretion disks with the dead zone is also discussed 
within the paradigm of protoplanetary disks (Nakano 1991; Jin 1996;
Gammie 1996; Sano \& Miyama 1999; Sano et al. 2000; Armitage et al. 2001). 
The origin of the dead zone in protoplanetary disks is the ohmic dissipation. 
Because protoplanetary disks are so cold that the ionization fraction
is very low. 
The ohmic dissipation of the electric current is efficient enough to
suppress the linear growth of the MRI.
In the next subsection, we discuss the typical evolutionary timescales
and energetics of the episodic accretion in collapsar disks referring
to the studies for protoplanetary disks. 

\subsection{Timescales and Energetics of the Episodic Accretion} 
We consider an ADAF-type disk ($p= 1.0,q=0.5$) described by
equations~(\ref{eq4})--(\ref{eq10}) as an example. Then the baryonic
mass of the inner dead zone is 
\begin{equation}
M_{\rm dead}  =  \int_{r_{\rm s}}^{r_{\rm crit}} 2\pi r \Sigma dr  
\simeq  0.15  f_\Sigma M_3^2 
\left( \frac{r_{\rm crit}}{20r_{\rm s}}  \right)^{3/2} M_{\sun} 
\;. \label{eq19}
\end{equation}
Thus, the most of the baryonic matter is stored 
in the outer active region at the early evolutionary stage.
Since Toomre's $Q$ value of the disk is
\begin{equation}
Q \equiv \frac{\Omega c_{\rm s}}{\pi G \Sigma} \simeq 7.8
f_{\Sigma}^{-1} f_T^{1/2}M_{3}^{-1}
\left( \frac{r_{\rm crit }}{20r_{s}}  \right)^{-3/2} \;, \label{eq20}
\end{equation}
the dead zone region is in the quiescent phase at the beginning.
As time passes, the matter from the outer active region is accumulated 
near the critical radius.
The gravitational instability begins to grow 
when the surface density at the critical radius becomes  
an order of magnitude larger than that of the initial quiescent phase (i.e., $f_{\Sigma } \gtrsim 8$). 

Once the outer edge of the dead zone becomes gravitationally unstable, 
spiral waves would be exited globally in the dead zone, and which
cause the intense mass accretion onto the central black hole.
The mass accretion rate due to the gravitational torque can be described as
\begin{equation}
\dot{M}_{\rm g} \simeq 0.35 f_{\Sigma} f_{\rm T} M_3 
\left(\frac{\alpha_{\rm g}}{0.05} \right) 
\dot{M}_\sun \;, \label{eq21} 
\end{equation}
where $\alpha_{\rm g} = 0.05$ is the $\alpha$-parameter in
gravitationally unstable disks (e.g., Lodato \& Rice 2005). 
The duration of the intense accretion phase should be comparable to 
the viscous timescale in the dead zone (Armitage et al. 2001);
\begin{equation}
\tau_{\rm g} = \frac{r^2}{\nu} \simeq 0.93 f_{\rm T}^{-1}M_3
\left( \frac{\alpha_{\rm g}}{0.05} \right) 
\left( \frac{r}{20r_{\rm s}} \right)^{3/2}\ \ {\rm sec} \;.\label{eq22}
\end{equation}
The gravitational energy released during an intense accretion phase is  
\begin{equation}
E_{\rm g} \simeq \eta \dot{M}_{\rm g} \tau_{\rm g} c^2  \simeq 2.4
\times 10^{53} f_{\Sigma} M_3^2 
\left( \frac{\alpha_g}{0.05} \right)^2
\left( \frac{r}{20r_{\rm s}} \right)^{3/2}
\left( \frac{\eta}{0.42}\right)\ {\rm erg}
\;, \label{eq23} 
\end{equation} 
where $\eta$ is the energy conversion efficiency from the rest-mass
energy to the gravitational one for the case of Kerr black holes
(Shapiro \& Teukolsky 1983). 
 
After the intense mass accretion phase, the dead zone returns to a
gravitationally stable state because of the decrease of the mass in
the dead zone region.
However, the mass accretion from the outer active region to the dead
zone continues constantly.
The quiescent disk progressively evolves to the intense
accretion phase again. 
This cycle would be repeated and the explosive energy release occurs
intermittently until the baryonic matter of the outer active region is
exhausted. 
Finally, the dead zone disk with a low mass accretion rate
would be left after the episodic accretion stage (see Fig.~\ref{fig4}$c$). 

If the episodic accretion is the origin of multiple relativistic
shells, which are expected to cause the internal shocks (Kobayashi et
al. 1997), the typical variable timescale $\tau_{\rm var}$ in the prompt
emission  of GRBs should correspond to the duration of an intense
accretion phase:
\begin{equation}
\tau_{\rm var} \simeq \tau_{\rm g} \simeq 0.93 \ \ {\rm sec} \;.
\label{eq24} 
\end{equation}
This could be the origin of the observable 
log-normal feature in the short-term variability of the prompt emission 
(Nakar \& Piran 2002; Kobayashi et al. 2002; Shen \& Song 2003). 
When a few percent of the gravitational energy is converted to the
radiative one, 
the typical luminosity is evaluated as 
\begin{equation}
L_{\rm var} \simeq f\ \frac{E_{\rm g}}{ \tau_{\rm var}} \simeq 2.6 \times  10^{51}
\left( \frac{f}{0.01} \right)\ \ {\rm erg\ sec^{-1}} \;, \label{eq25} 
\end{equation}
where $f$ is the conversion factor from the gravitational energy to
the radiative one. 
These are almost identical to the observed timescale and peak
luminosity of variable components in the prompt emission (Norris et
al. 1996; Nakar \& Piran 2002). 

The episodic mass accretion is terminated when the material in the
outer active region is exhausted. 
Then, the total duration of the prompt emission will be
determined by the mass depletion timescale.
Applying the mass inflow rate given by equation~(\ref{eq18}), 
we can evaluate the total duration and luminosity of the prompt
emission as 
\begin{equation}
\tau_{\rm tot}  \simeq  M_{\rm tot}/ \dot{M}_{\rm in} \simeq
\mathcal{O} (10) \ \ {\rm sec} \;, \label{eq26} \\ 
\end{equation}
\begin{equation}
L_{\rm tot}  \simeq  f\ \frac{\eta M_{\rm tot}c^2}{ \tau_{\rm tot}}
\simeq \mathcal{O}(10^{51}) \ \ {\rm erg\ sec^{-1}} \;, \label{eq27}
\end{equation}
where $M_{\rm tot}$ is the total mass of the accretion disk and is
assumed to be 1 -- 2 $M_\sun$. Thus, our episodic accretion model 
of the collapsar disk can explain many observed features of long GRBs quantitatively. 

\section{DISCUSSION}
Recent {\it{Swift}} observations have discovered the flaring activity
in the early X-ray afterglow occurring $\approx 10^3$ -- $10^4$ sec
after the prompt burst
(Burrows et al. 2005; Nousek et al. 2006; O'Brien et al. 2006). 
The flare is suggested to be a prolonged activity of 
the central engine, and various models are proposed for 
explaining its origin (Ioka et al. 2005; King et al. 2005; 
Proga \& Zhang 2006; Perna et al. 2006). 
How can the flaring activities be interpreted 
in the context of our episodic accretion scenario?

We consider the case that the outermost region of the collapsar disk
is gravitationally unstable, that is, $Q \lesssim 1$ around $70 r_{\rm s}$. 
When the disk is gravitationally unstable, it evolves toward the
spirally-undulated structure or fragments into bound 
objects. Fragmentation occurs when the local cooling time in the disk 
is shorter than the orbital time $\Omega_{\rm K}^{-1}$ (Gammie 
2001; Rice et al. 2003). In this case, the local cooling 
dominates the viscous heating sustained by the gravitational instability. 

In the outermost region of the collapsar disk, the cooling of 
$^4\rm{He}$ photodisintegration can be effective 
(Piro \& Pfahl 2006; Kawanaka \& Mineshige 2007). 
The cooling time would be much smaller than the orbital time there.
Thus, if the outer region of collapsar disk is gravitationally
unstable, it can fragment into multiple bound objects. 
The mass of a bound fragment is estimated as $m_f \sim 
0.2 M_{\sun } $ assuming the coalescence of small fragments 
into a single body (Piro \& Pfahl 2006). 

Considering that the orbital angular momentum of the fragment is
mainly extracted in the form of gravitational radiation, its infalling time is typically
\begin{equation}
\tau_{\rm gw}  \simeq 6.5 \times 10^{3} M_3^2 
\left( \frac{m_f }{0.2 M_{\sun}} \right)^{-1}
\left( \frac{r}{70r_s } \right)^4\ {\rm sec} \ \ \ \;, \label{eq29}
\end{equation}
(Peters 1964; Piro \& Pfahl 2006), which agrees with the typical
arrival time of the X-ray flares (Burrows et al. 2005; Nousek et
al. 2005; O'Brien et al. 2006). 
Note that this is much longer than the evolutionary timescale of
hyperaccretion disks discussed in the previous section.

As the fragment approaches to the central black hole, 
it suffers strong tidal force. 
Then, the bound fragment is disrupted and a rotating disk 
of its mass $\sim 0.2 M_\sun$ is formed around the black hole.
The mass of this newly formed disk (remnant disk) is comparable to the
mass of the dead zone region [see eq.~(\ref{eq19})].
Therefore, the entire region of the remnant disk is magnetorotationally
inactive if the field strength is 
weaker than the critical value, $B_{\rm crit} \approx 10^{14}\ \rm{G}$. 
Unlike the early evolutionary stage of the collapsar disk, 
the episodic accretion can not be expected in the remnant disk. 
This is because the baryonic matter does not exist at the outside of
the dead zone. Therefore, the evolutionary timescale is determined by the neutrino
viscosity ($\alpha_{\nu} \sim 10^{-4}$) and the mass accretion rate is 
relatively low, $\dot{M}_{\rm out} 
\simeq 7.2 \times 10^{-4} \dot{M}_{\sun}$ [c.f., eq(\ref{eq17})]. 
The duration of the stationary mass accretion would be
\begin{equation}
\tau_{\rm rem}  \simeq  m_f/ \dot{M}_{\rm out} \simeq \mathcal{O} (100)  \ \ \ {\rm sec} \;, \label{eq30}
\end{equation}
which is consistent with the observed duration of the X-ray flares
(Chincarini et al. 2007). 

\section{SUMMARY}
In this paper, we have investigated where the MRI operates in
hyperaccretion disks focusing on the effect of the neutrino viscosity 
in order to reveal the energy releasing mechanism of GRBs. 
Our main findings are summarized as follows:

1. Assuming reasonable disk models and simple neutrino transport, 
the MRI is suppressed by the neutrino viscosity in the inner
neutrino-opaque region of hyperaccretion disks when the magnetic field
is weaker than the critical value, $B_{\rm crit} \simeq 10^{14}{\rm G}$. 
On the other hand, in the outer neutrino-transparent region, 
the interaction between the neutrino and baryonic matter can be neglected, 
and thus the MRI can drive MHD turbulence actively. 
Most of the disk models show that hyperaccretion disks consist of
the inner dead zones and outer active regions.

2. If the MRI is suppressed in the dead zone, the baryonic matter
accreted from the outer active region should accumulate there. 
The mass inflow rate from the active region is estimated as 
$\dot{M}_{\rm in} \approx 10^{-2}\dot{M}_\sun$. 
Then, the inner part of the disk may become gravitationally unstable 
at some evolutionary stage. This suggests that intense mass accretion
onto the central black hole could be caused episodically 
by the gravitational torque. 

3. The typical evolutionary timescale of the intense accretion phase 
corresponds to the viscous timescale of the gravitationally unstable
dead zone, 
and is evaluated as $\tau_{\rm g} \approx 1$ sec. If the episodic
release of the gravitational 
energy is the origin of multiple relativistic shells, the typical
variable timescale and peak luminosity 
in the prompt burst are given by $\tau_{\rm var} \approx \tau_{\rm
  g} \simeq \mathcal{O}(1)$ sec and 
$L_{\rm var} \simeq \mathcal{O}(10^{51})\ {\rm erg\ sec^{-1}}$. These
are consistent with the observational features of the short-term
variability in GRBs. 

4. The total duration of the prompt burst should be determined by
the mass depletion timescale of the outer active region, 
because the episodic mass accretion is
terminated when the material in the active region is
exhausted. Considering a collapsar disk of mass $\sim 1$ -- 2
$M_{\sun }$, we can evaluate the total duration of the prompt burst
as $\tau_{\rm tot} \simeq \mathcal{O}(10)$ sec. 
In addition, the typical luminosity of the prompt burst is obtained
as $L_{\rm tot} \simeq \mathcal{O}(10^{51})\ {\rm erg\ sec^{-1}}$ in
our episodic accretion model. 

\acknowledgments
We thank N. Turner for his careful reading of the manuscript. 
We thank S. Inutsuka, K. Ioka, K. Toma, T. Nakamura, H. Kamaya, T. Takiwaki,
T. Kawamiti and Y. Mamiduka for helpful discussions. 
TS is supported by the Grant-in-Aid (16740111, 17039005) from the 
Ministry of Education, Culture, Sports, Science, and Technology of Japan. 
This work is supported by the Grant-in-Aid for the 21st Century COE 
``Center for Diversity and Universality in Physics'' from the Ministry
of Education, Culture, Sports, Science and Technology of Japan. 



\clearpage
\clearpage
\clearpage
\clearpage 
\clearpage
\end{document}